\documentclass[12pt]{article}
\usepackage{a4wide}
\usepackage{bbold}
 \usepackage{pstricks}
\usepackage{amsmath,amssymb,bbm}
\usepackage{slashed}
\usepackage{multirow}
 \usepackage{graphicx}
 \usepackage[verbose]{wrapfig}
\usepackage{comment}
\newsavebox{\uuunit}
\sbox{\uuunit}
    {\setlength{\unitlength}{0.825em}
     \begin{picture}(0.6,0.7)
        \thinlines
        \put(0,0){\line(1,0){0.5}}
        \put(0.15,0){\line(0,1){0.7}}
        \put(0.35,0){\line(0,1){0.8}}
       \multiput(0.3,0.8)(-0.04,-0.02){12}{\rule{0.5pt}{0.5pt}}
     \end {picture}}

\def\2{\frac12}
\def\4{\frac14}

\newcommand{\be}{\begin{equation}}
\newcommand{\ee}{\end{equation}}
\newcommand{\bea}{\begin{eqnarray}}
\newcommand{\eea}{\end{eqnarray}}

\usepackage{hyperref}
\hypersetup{
linkbordercolor={blue}
}

\hypersetup{
    bookmarks=false,         % show bookmarks bar?
    unicode=false,          % non-Latin characters in Acrobat’s bookmarks
    pdftoolbar=true,        % show Acrobat’s toolbar?
    pdfmenubar=true,        % show Acrobat’s menu?
    pdffitwindow=false, % window fit to page when opened
    pdfstartview={FitH},    % fits the width of the page to the window
    pdftitle={Non-Geometric Fluxes from Exotic Branes},    % title
    pdfauthor={Eric A.Bergshoeff,Tomas Ort\'\i n,Fabio Riccioni},     % author
    pdfsubject={},   % subject of the document
    pdfnewwindow=true,      % links in new window
    colorlinks=false,       % false: boxed links; true: colored links
    linkcolor=blue,          % color of internal links
    citecolor=blue,        % color of links to bibliography
    filecolor=blue,      % color of file links
    urlcolor=blue           % color of external links
    linkbordercolor={blue},
    citebordercolor={purple},
    urlbordercolor={gray}
}

\def\equationautorefname~#1\null{%
  eq.~(#1)\null
}

\def\figureautorefname~#1\null{%
  Fig.~#1\null
}

\begin{document}

\begin{titlepage}%1
\begin{center}

\hfill UG-15-97 \\

\vskip 1.5cm

{\Large \bf Non-geometric fluxes
and  \vskip .7cm
mixed-symmetry potentials}

\vskip 1.5cm

{\bf  E.A.~Bergshoeff\,$^1$,  V.A. Penas\,$^1$, F. Riccioni\,$^2$  and S. Risoli\,$^3$}

\vskip 30pt

{\em $^1$ \hskip -.1truecm Centre for Theoretical Physics,
University of Groningen, \\ Nijenborgh 4, 9747 AG Groningen, The
Netherlands \vskip 5pt }

\vskip 15pt

{\em $^2$ \hskip -.1truecm
 INFN Sezione di Roma,   Dipartimento di Fisica, Universit\`a di Roma ``La Sapienza'',\\ Piazzale Aldo Moro 2, 00185 Roma, Italy
 \vskip 5pt }

\vskip 15pt

{\em $^3$ \hskip -.1truecm  Dipartimento di Fisica and INFN Sezione di Roma, Universit\`a di Roma ``La Sapienza'',\\ Piazzale Aldo Moro 2, 00185 Roma, Italy
 \vskip 5pt }

\end{center}

\vskip 0.5cm

\begin{center} {\bf ABSTRACT}\\[3ex]
\end{center}

We discuss the relation between generalised fluxes and mixed-symmetry potentials. We first consider the NS fluxes, and point out that the `non-geometric' $R$ flux is dual to a mixed-symmetry potential with a set of nine antisymmetric indices.
We then consider the T-duality family of fluxes whose prototype is  the Scherk-Schwarz reduction of the S-dual of the RR scalar of IIB supergravity. Using the relation with mixed-symmetry potentials, we are able to give a complete classification of these fluxes, including the ones that are non-geometric. The non-geometric fluxes again turn out to be dual to potentials containing nine antisymmetric indices. Our analysis suggests that
all these fluxes can be understood in the context of double field theory, although for the non-geometric ones one expects a violation of the strong constraint.

\end{titlepage}

\newpage
\setcounter{page}{1} \tableofcontents

%\newpage
\vskip 2truecm

%%%%%%%%%%%%%%%%%%%%%%%%%%%%%%%%%%%%%%%%%%%%%%%%%%%%%%%%%%%%%
\setcounter{page}{1} \numberwithin{equation}{section}

\section{Introduction}

Supergravity theories arise as the low-energy effective actions of string theories. In general, when one considers string theory on a background that preserves some amount of supersymmetry, the resulting supergravity theory contains moduli, {\it i.e.}~scalars that are not stabilised by any potential. On the other hand, it is of phenomenological interest to construct models in which such moduli are stabilised. This is in general achieved by introducing fluxes, which means that some field-strengths with indices in the internal directions have a non-trivial background value. The fluxes induce a gauging in the lower-dimensional supergravity theory, which is indeed related by supersymmetry to a potential that may stabilise  the scalars.

String theories are conjectured to be related by discrete non-perturbative dualities. It is natural to ask what happens when one performs such dualities in the presence of fluxes. One thing that can happen in particular is that a flux is mapped by duality to a `non-geometric' flux, that is something that cannot be obtained in terms of the fields of the higher-dimensional supergravity theory. The non-geometric nature of these fluxes mimics the fact that the string dualities themselves cannot be simply understood in terms of geometric isometries. From the point of view of the low-energy effective action, string dualities appear as continuous global symmetries of the supergravity theory. The way a gauging is mapped to another gauging by duality is encoded in the so-called `embedding tensor' \cite{Nicolai:2000sc}, which means that the constant parameter that identifies the gauging can be formally considered as a tensor of the global symmetry group. This paper is concerned with  maximal supergravity theories, and all the possible embedding tensors of these theories have been classified \cite{embeddingtensorreprs}.

In this paper we are focused on how fluxes transform under T-duality, which is a perturbative duality symmetry of string theory  exchanging momenta and winding modes of the string. In the case of maximal theories in $10-d$ dimensions, this symmetry is $\text{O}(d,d;\mathbb{Z})$. In particular, one can consider how T-duality acts on the Ramond-Ramond (RR) fluxes. The simplest example of such fluxes is the Scherk-Schwarz (SS) reduction of IIB supergravity to nine dimensions, in which the RR scalar $C$ acquires a linear dependence on the internal coordinate $x^9$, {\it i.e.}~$C= C(x^m )+ M x^9$, where $M$ is a constant and $m =0,...,8$. This ansatz leads to a consistent truncation to $D=9$, because $C$ only occurs in the IIB action via derivatives, and the resulting nine-dimensional theory is a gauged supergravity. In nine dimensions, the only T-duality symmetry is the one exchanging IIA and IIB supergravity. Therefore, one expects that what T-duality does in this case is to provide a IIA supergravity origin for the same nine-dimensional gauging. This is just the dimensional reduction of Romans' massive IIA theory \cite{Romans:1985tz}.

In lower dimensions $D=10-d$ with $d>1$ one can have more general RR fluxes by giving a constant vacuum expectation value to any of the RR $p$-form field strengths $G_{p}$, provided that $p$ is less or equal to $d$. Considering a democratic formulation, in which both the electric and magnetic RR potentials are introduced, $G_{p}$ can be any even form in IIA and any odd form in IIB supergravity. In the absence of fluxes, the low-energy supergravity theory possesses a global $\text{SO}(d,d)$ symmetry, and the RR gaugings are identified by an embedding tensor $\theta_{\alpha}$, which is a chiral spinor of $\text{SO}(d,d)$.\,\footnote{The symmetry in the NS sector is $\text{O}(d,d)$. Note that there is no chirality in $\text{O}(d,d)$.}
Such spinor has $2^{d-1}$ components, and it decomposes in even-rank  or odd-rank antisymmetric representations  of $\text{SL}(d,\mathbb{R})$ according to the convention chosen for its chirality: denoting with ${{\bf d} \choose n}$ the $\text{SL}(d,\mathbb{R})$ antisymmetric representation with $n$ downstairs indices,\footnote{Similarly, in the paper we denote with ${{\bf \overline{d}} \choose n}$ the antisymmetric representation with $n$ upstairs indices.}, one gets
\begin{eqnarray}
& & ( 2^{d-1} )_{\rm S} = {\bf 1} \oplus {{\bf d }\choose {2}} \oplus {{\bf d }\choose { 4}} \oplus ...\quad ,\label{alphaindexIIAconvention}
\\
& & ( 2^{d-1} )_{\rm C} = {\bf d} \oplus {{\bf d }\choose {3}} \oplus {{\bf d }\choose {5}} \oplus ... \quad ,\label{alphaindexIIBconvention}
\end{eqnarray}
where the two chiral-spinor representations of $\text{SO}(d,d)$ are identified by the index ${\rm S}$ and ${\rm C}$ as usual.  The first equation corresponds to the decomposition in terms of the IIA fluxes and the second to the one in terms of the IIB fluxes.
Each term on the right-hand side of the above two equations corresponds to a
geometric flux $G_{a_1 ...a_p} = \partial_{[a_1} C_{a_2 ...a_p ]}$, where $C_{a_1 ...a_{p-1}}$ is a 10-dimensonal RR potential. The only exception is the singlet in \autoref{alphaindexIIAconvention}, which corresponds to the dimensional reduction of the mass parameter of Romans' IIA supergravity. As \autoref{alphaindexIIAconvention} and \autoref{alphaindexIIBconvention} show, going from IIA to IIB supergravity corresponds to changing the convention for the chirality property of $\theta_\alpha$.  We list in \autoref{RRgaugings} the explicit decomposition of \autoref{alphaindexIIAconvention} and \autoref{alphaindexIIBconvention} in each dimension.

\begin{table}[h!]
\renewcommand{\arraystretch}{1.3}
\par
\begin{center}
\scalebox{1.1}{
\begin{tabular}{|c||c||c|c|c|c||c|c|c|}
\hline
$D$ & $\theta_\alpha$ & $G$  & $G_{a_1 a_2}$ & $G_{a_1 ... a_4}$ & $G_{a_1 ...a_6}$ & $G_a$ & $G_{a_1 a_2 a_3}$ & $G_{a_1 ... a_5}$ \\
\hline
\hline 9 & ${ 1}$ & ${\bf 1}$ & &&  & ${\bf 1}$ & &
 \\
 \hline
8 & ${2}$ & ${\bf 1}$ & ${\bf 1}$  && & ${\bf 2}$ & & \\
\hline
7 & ${4}$ &   ${\bf 1}$ & ${\bf \overline{3}}$  && &   ${\bf 3}$ & ${\bf 1}$  & \\
\hline
6 & ${8}$ &   ${\bf 1}$ & ${\bf 6}$  & ${\bf 1}$ &   &   ${\bf 4}$ & ${\bf \overline{4}}$  &  \\
\hline
5 & ${16}$ &   ${\bf 1}$ & ${\bf 10}$  & ${\bf \overline{5}}$ & &   ${\bf 5}$ & ${\bf \overline{10}}$  & ${\bf 1}$   \\ \hline
4 & ${32}$ &   ${\bf 1}$ & ${\bf 15}$  & ${\bf \overline{15}}$ & ${\bf 1}$ &   ${\bf 6}$ & ${\bf 20}$  & ${\bf \overline{6}}$  \\
\hline
\end{tabular}
}
\end{center}
\caption{\footnotesize The decomposition of the RR embedding tensor $\theta_\alpha$, which is a chiral spinor of $\text{SO}(d,d)$, in terms of the RR fluxes of the IIA
(left of vertical double line in the middle) and IIB theory (right of vertical double line in the middle), written as representations of $\text{SL}(d,\mathbb{R})$ in $D=10-d$ dimensions.}
\label{RRgaugings}
\end{table}

A similar analysis can be performed for the so-called NS fluxes, which are those related by T-duality to the NS 3-form flux $H_{abc}$, with $a=1,\dots,d$. In any dimension, the embedding tensor arising from these fluxes is $\theta_{MNP}$, with $M=1,\dots ,2d$, belonging to the three-index completely antisymmetric representation of $\text{SO}(d,d)$, which decomposes in terms of $\text{SL}(d,\mathbb{R})$ representations to give the well-known chain of NS fluxes \cite{Shelton:2005cf}
\begin{equation}
\theta_{MNP} \ \rightarrow \ H_{abc} \quad f_{ab}{}^c \quad Q_{a}{}^{bc} \quad R^{abc} \quad . \label{listofNSfluxes}
\end{equation}
In this equation, $f_{ab}{}^c$ denotes the metric flux, which is the only geometric flux apart from $H_{abc}$. The $Q_a{}^{bc}$ and $R^{abc}$ fluxes are both  non-geometric, but their non-geometric nature is very different: while the former can be written as $Q_a{}^{bc} = \partial_a \beta^{bc}$, which is a SS reduction of a suitable combination $\beta^{ab}$ of the NS 2-form $B_{ab}$ and the metric (and hence it is dubbed `locally geometric'), there is no possible geometric interpretation for the latter within supergravity. This is similar to what happens to the RR flux in nine dimensions, which cannot be obtained geometrically from the IIA massless theory. The difference is that in the case of the RR flux, as discussed above, the T-dual origin of the flux arises in terms of a deformation of the ten-dimensional IIA theory, which is the massive IIA theory of Romans. However, there is no equivalent to the Romans theory in the NS sector. Hence, it is impossible to have a higher-dimensional origin of a purely $R^{abc}$ flux within supergravity.

In Double Field Theory (DFT) \cite{Siegel,DFTBasics}  the non-geometric nature of the $R$ flux manifests itself in a very clear way. In the context we are discussing, DFT provides a fully $\text{O}(d,d)$-covariant way of obtaining the NS fluxes by doubling the internal coordinates and hence the $R$ flux can be written as a SS reduction in doubled space, {\it i.e.}~$R^{abc} = \tilde{\partial}^{[a}  \beta^{bc ]}$. Supergravity is recovered from DFT by imposing the so-called strong constraint, that forces all the fields to depend only on half of the coordinates. If one chooses the geometric coordinates to be the $x$'s, clearly having an ansatz in which a field depends on $\tilde{x}$ violates the strong constraint. Actually, also the action of T-duality on the nine-dimensional RR gauging can be understood in DFT \cite{Hohm:2011cp}. Indeed, the ansatz $C= C(x^m ) + M x^9$, with $m = 0,...,8$, where $C$ is the RR scalar of the IIB theory, is mapped by T-duality to\,\footnote{This should not be confused with the 9-form $C_9$ which is introduced in the next paragraph.} $C_9 = C_9 ( x^m ) + M \tilde{x}_9$, where $C_9$ is the component of the RR 1-form $C_\mu$ along the internal direction. In this case, though, because of the way the RR fields are described in DFT \cite{Hohm:2011zr} and the linear dependence on $\tilde{x}_9$, this violation of the strong constraint is still consistent and precisely leads to the Romans theory \cite{Hohm:2011cp}.
% On the contrary, violating the strong constraint is not consistent in the NS sector, which is the DFT equivalent of the statement that there is no higher-dimensional supergravity origin for the gauged supergravity theory that one obtains turning on the $R$ flux.

The analysis of \cite{Hohm:2011cp} shows that the Romans mass parameter can be thought as the 0-form field strength $G_0$ of the 1-form $C_1$ in doubled space, {\it i.e.}~$G_0 = \tilde{\partial}^\mu C_\mu$. On the other hand,
the democratic formulation of the RR fields implies that in IIA supergravity one can introduce a 9-form $C_9$ whose field strength $G_{10}$ is the Hodge dual of the Romans mass parameter $G_0$. The special thing about this duality relation is that it maps a non-geometric configuration for the 1-form $C_1$ to a fully geometric configuration for $C_9$.
In general, in any dimension $D$ one can introduce $D-1$-form potentials which are dual to the embedding tensor.   It  can be  shown that all such $D-1$ forms in maximal supergravity theories are obtained from the mixed-symmetry potentials that arise in the decomposition of the adjoint representation of $\text{E}_{11}$ \cite{West:2001as} corresponding to the ten-dimensional IIA and IIB theories  \cite{Riccioni:2007au}. These mixed-symmetry potentials can be divided into three sets:
\begin{itemize}

\item the actual fields of the ten-dimensional theory, that are the metric, the scalars and all the forms (electric and magnetic duals), together with the `dual graviton', which is a mixed-symmetry potential in the (7,1) Young-Tableaux representation;

\item  mixed-symmetry potentials with one set of eight antisymmetric indices, {\it i.e.} in (8,...) Young Tableaux representations;

\item mixed-symmetry potentials with one set of nine antisymmetric indices (the RR 9-form $C_9$ is a special case in this set, because it has nine antisymmetric indices but it is not a mixed-symmetry potential).

\end{itemize}

The full list of mixed-symmetry potentials that give rise to the $D-1$ form dual to the NS embedding tensor $\theta_{MNP}$ was given in \cite{Bergshoeff:2011zk}. In this paper we first want to expand in this direction. In particular, we will show that

\begin{itemize}

\item

the geometric fluxes $H$ and $f$ are dual to potentials belonging to the first set;

\item

the locally geometric flux $Q$ is dual to a potential belonging to the second set;

\item

the non-geometric flux $R$ is dual to a potential belonging to the third set.

\end{itemize}
The first correspondence between fluxes and mixed-symmetry potentials is straightforward, see also the next section.  In order to understand the second correspondence, one can use the observation \cite{Riccioni:2006az} that the mixed-symmetry fields in the second set can be thought of as generalised duals of the standard supergravity fields \cite{deMedeiros:2002ge,ChatGau}. Therefore,  they do not correspond to new fields and one can expect that they are dual to redefinitions of the supergravity fields depending on the standard coordinates. The mixed-symmetry fields in the third set are instead fields that do not satisfy any generalised duality relation in ten dimensions, they arise as deformation parameters only when they are reduced to $D-1$ forms. In this sense, the RR 9-form is an exception because it is already a form in ten dimensions, which is the dual counterpart of the statement that the violation of the strong constraint in the RR sector discussed in \cite{Hohm:2011cp} is still consistent within DFT. To summarise, the main result of our analysis is that mixed-symmetry `dual' potentials in normal space are equivalent to standard potentials in double space
with a possible nontrivial dependence on $\tilde{x}_\mu$. Furthermore,  the inconsistency of the
standard potentials in double space  due to the violation of the strong constraint is equivalent to the impossibility of describing mixed-symmetry potentials in supergravity
in a consistent way.

The main result of this paper will be the generalisation of this analysis to another family of fluxes, which are those related by T-duality to the IIB SS reduction of the S-dual of the axion $C$. As we will discuss, the embedding tensor for the resulting gaugings belongs to the `gravitino' irreducible representation of $\text{SO}(d,d)$, which is $\theta_{M\alpha}$ for $d$ odd and $\theta_{M \dot{\alpha}}$ for $d$ even (the conventions for the spinor indices are fixed by denoting with $\theta_\alpha$ the embedding tensor for the RR gaugings in any dimension). The mixed-symmetry potentials
that generate by dimensional reduction the $D-1$ form potentials dual to this embedding tensor have been listed in \cite{Bergshoeff:2011ee}. We will show in this paper that the correspondence, discussed above for the NS sector, between locally geometric and non-geometric fluxes on the one side and mixed-symmetry potentials with eight and nine antisymmetric indices on the other side still holds. In particular, this implies  that all these new gaugings can in principle be described in the context of DFT precisely as the NS fluxes.

The plan of the paper is as follows. In \autoref{NSfluxessection} we first review the NS fluxes. Furthermore, we  discuss the duality with mixed-symmetry potentials. In \autoref{newfluxessection} we repeat the same analysis for the family of fluxes to which the SS reduction of the S-dual of the IIB axion belongs.
% In \autoref{dualityandbranessection} we discuss how the duality relations discussed in the paper could admit % a possible unifying description in DFT and speculate on how these duality relations can explain the
% occurrence of the so-called `wrapping rules' satisfied by the branes that couple to the potentials involved.
Finally, \autoref{conclusionssection} contains our conclusions.

\section{\label{NSfluxessection}NS fluxes revisited}

In this section we consider the NS fluxes, that are the fluxes related by T-duality to the $H_{abc}$ 3-form flux. We want to show that all these fluxes can be classified in terms of their dual fields, that are mixed-symmetry potentials in 10 dimensions. In particular, we will show that the non-geometric fluxes are dual to a mixed-symmetry potential that contains a set of nine antisymmetric indices.

The general classification of gaugings of maximal supergravity theories in any dimension in terms of the embedding tensor reveals that the gaugings resulting from turning on NS fluxes belong to the embedding tensor $\theta_{MNP}$ in the completely antisymmetric three-index representation of $\text{O}(d,d)$. Decomposing this representation in terms of representations of $\text{SL}(d, \mathbb{R})$ representations according to
\begin{equation}
({\bf 2d} \otimes {\bf 2d} \otimes {\bf 2d} )_A = ({\bf d} \otimes {\bf d} \otimes {\bf d} )_A \oplus [({\bf d} \otimes {\bf d} )_A  \otimes {\bf \overline{d}}]\oplus [{\bf {d}} \otimes ({\bf \overline{d}} \otimes {\bf \overline{d}} )_A ]\oplus [({\bf \overline{d}} \otimes {\bf \overline{d}} \otimes {\bf  \overline{d}} )_A
\end{equation}
one obtains the well-known T-dual family of fluxes  in \autoref{listofNSfluxes}
containing, together with the 3-form flux $H_{abc}$ and the metric flux $f_{ab}{}^c$, the two generalised fluxes $Q_a{}^{bc}$ and $R^{abc}$ \cite{Shelton:2005cf}.
In seven dimensions, the embedding tensor $\theta_{MNP}$ belongs to the ${\bf 10} \oplus {\bf \overline{10}}$ of $\text{SO}(3,3)$, where the reducibility of the representation is due to the splitting in selfdual and anti-selfdual part. The fluxes $H_{abc}$ and $Q_{a}{}^{bc}$ belong to the   ${\bf 10}$, while the fluxes $f_{ab}{}^c$ and $R^{abc}$ belong to the  ${\bf \overline{10}}$. Performing a single T-duality corresponds to swapping these two representations.

Starting from $D=10$ the $H_{abc}$ flux arises for the first time in seven dimensions ($H_{789}=\partial_{7}B_{89}$). The field $B_{89}$ has a linear dependence on $x^7$ and the flux can be seen as a SS reduction from $D=8$ to $D=7$ along the $x^7$ coordinate. By performing a T-duality along, say, $x^9$, the flux is mapped to $f_{78}{}^9$ as \autoref{listofNSfluxes} shows. This comes from a SS reduction of the metric components. From the $D=10$ point of view the background fields ($g_{\mu\nu},B_{\mu\nu,}\phi$) are related by the well known Buscher-rules.  When T-dualising along the isometry direction $x$ (the remaining directions are indicated by $i$) these rules read as follows:

\begin{equation}
g'_{ij}=g_{ij}-\frac{1}{g_{xx}}\left(g_{i x}g_{j x}-B_{i x}B_{j x}\right)\,,
\end{equation}

\begin{equation}
B'_{ij}=B_{ij}+\frac{2}{g_{xx}}g_{[i|x|}B_{j]x}\,,
\end{equation}

\begin{equation}
g'_{i x}=-\frac{B_{i x}}{g_{xx}}\qquad B'_{i x}=-\frac{g_{i x}}{g_{xx}} \qquad g'_{xx}=\frac{1}{g_{xx}}\,,
\end{equation}

\begin{equation}
\phi'=\phi -\frac{1}{2}\text{ln}|g_{xx}|  \ \ .
\end{equation}

If one performs a further T-duality, say along $x^8$, this leads to a $Q_7{}^{89}$ flux, which arises as a SS reduction for the ten dimensional field $\beta^{\mu\nu}$ which is defined in $\beta$-supergravity  \cite{betasupergravity} as follows:
\begin{equation}
\beta^{\mu\nu}=-((g-Bg^{(-1)}B)^{-1})^{\mu\sigma}B_{\sigma\rho}g^{\rho\nu} \quad .
\end{equation}
In particular, in $D=8$ this gives
\begin{equation}
\beta^{89} = - \frac{B_{89}}{{\rm det}g + (B_{89} )^2 } \label{betaeightnine} \quad ,
\end{equation}
where ${\rm det} g$ is the determinant of the metric in the 8 and 9 directions.
Defining the complex scalar $\rho= B_{89} + i \sqrt{{\rm det} g}$, that parametrises the  $\text{SL}(2,\mathbb{R})/\text{SO}(2)$ part of the scalar isometry that transforms the $B$ field, performing two T-dualities along the 8 and 9 directions leads to the transformation $\rho \rightarrow - 1/\rho$, in agreement with the fact that $\beta^{89}$ in \autoref{betaeightnine} is the real part of $-1/\rho$.

Although T-duality implies the presence of the $R^{789}$ flux, performing a further T-duality along $x^7$ is problematic because the field $\beta^{89}$ has a linear dependence on $x^7$. This is the reason why this flux is dubbed purely `non-geometric'. As discussed in the introduction, in the RR case one encounters a similar problem when one wants to understand from the IIA perspective the 1-form flux corresponding to a SS reduction of the IIB axion. The difference is that in that case Romans' massive IIA supergravity  \cite{Romans:1985tz} precisely provides this T-dual origin. In the case of the $R^{789}$ flux, instead, such a massive supergravity theory in dimension higher than seven does not exist.

One can understand the same non-geometric properties as arising by considering the branes that are sources for these fluxes. In particular, the brane that sources the $Q_7{}^{89}=\partial_7\beta^{89}$ flux is the so-called $5_2^2$-brane smeared along the $x_7$ direction. This brane is known as a T-fold  since when one circles around the brane in transverse space the metric does not come back to the same point \cite{deBoer:2010ud}. The nontrivial monodromy can be understood as a shift in the $\beta$-field, that in the $5_2^2$ background takes the simple form
\begin{equation}
\beta^{89} =- \frac{B_{89}}{g_{88}g_{99}+(B_{89})^2} \quad .
\end{equation}
When smearing along $x^7$ one obtains a harmonic function that is linear in the only remaining transverse direction, while for consistency the field $\beta^{89}$ must acquire a linear dependence on $x^7$, exactly as in the D7-brane case. As before, the question is how can one perform a T-duality along $x^7$.

Double field theory (DFT) provides an approach to deal with this issue. In DFT, all fields depend on $X^M=(x^\mu,\tilde{x}_\mu)$ where $x^\mu$ are the usual space-time coordinates and $\tilde{x}_\mu$ are the winding coordinates. The theory is equipped with the so-called {\it strong constraint}, {\it i.e.}~any combination of fields $A$, $B$ must satisfy
\begin{equation}
\partial_M\partial^M A=0 \ \ ,~~~~ \partial_MA\partial^MB=0 \quad .
\end{equation}
These constraints imply that one can always rotate to a frame where the fields depend on half of the coordinates. In DFT, T-duality swaps $x$ and $\tilde{x}$, which implies that the SS ansatz $\beta^{89} = \beta^{89} (x) + Q_7{}^{89} x^7$ corresponding to the $Q$-flux is mapped to $\tilde{\beta}^{89} = \tilde{\beta}^{89} (x) + R^{789} \tilde{x}_7$. In the  supergravity frame, {\it i.e.}~the frame where all the fields depend on $x$ only, the $Q$-flux ansatz satisfies the strong constraint. But after performing a T-duality to obtain the $R$-flux, the dual background necessarily will depend on a dual coordinate, thus violating the supergravity frame. The dual coordinate dependence in the $R$-flux ansatz is actually compatible with a generalized SS reduction of DFT, in the sense that reductions on both standard and dual internal coordinates are allowed \cite{GaugedDFT}. Exactly the same applies for the corresponding domain-wall solutions: if one performs a T-duality along $x^7$ on the smeared $5_2^2$-brane solution discussed above in DFT, one obtains a so-called $R$5-brane, which is a domain wall in seven dimensions with $\beta$ depending linearly on $\tilde{x}^7$.

This is analogous to what happens in the RR sector. In that case the ansatz $C = C(x) + m x^9$ is mapped to $C_\mu = C_{\mu} (x) + \delta_{\mu 9} m \tilde{x}_9$ \cite{Hohm:2011cp} when one performs a T-duality along $x^9$. The difference with the previous case is that in the case of the RR sector the violation of the strong constraint leads to a well-defined ten-dimensional theory, which is Romans' massive IIA supergravity theory \cite{Hohm:2011cp}. In the case of the NS sector, instead, such a violation will not lead to a consistent theory in ten dimension (or in nine and eight, for that matter). This result is the DFT equivalent of the statement that in the case of the RR fluxes the massive deformation  corresponds to  a massive theory in ten dimensions, while in the case of the NS fluxes such a massive theory does not exist in dimension higher than seven.

As we mentioned in the introduction, it is well known that one can consider the mass of the Romans theory as the dual of the 10-form field strength of the 9-form RR potential $C_9$. Similarly, the embedding tensor $\theta_{MNP}$ is dual in any dimension $D$ to a $D-1$-form potential $D_{D-1, MNP}$.
Starting from $d\ge 3$ or, equivalently, $D\le 7$, the duality relation (neglecting the contribution from any other field) has the schematic form
\begin{equation}
\frac{1}{\sqrt{|g|}}\epsilon^{\mu_1 ...\mu_D} \partial_{\mu_1} D_{\mu_2 ..\mu_D, MNP} = {\cal M}_M{}^Q {\cal M}_N{}^R {\cal M}_P{}^S\theta_{QRS} \quad , \label{dualityNSpotentialsDdim}
\end{equation}
where ${\cal M}$ parametrises the coset $\text{O}(d,d)/[\text{O}(d)\times \text{O}(d)]$ and can be thought of as the DFT generalised metric ${\cal H}$ of $\text{O}(d,d)$ with $G$ and $B$ only dependent on the $D$-dimensional spacetime coordinates. These ${\cal M}$'s are needed to have a duality relation that transforms covariantly
under $\text{O}(d,d)$.

The field $D_{D-1,MNP}$ arises from the dimensional reduction of the ten-dimensional mixed-symmetry fields \cite{Bergshoeff:2011zk}\,\footnote{In these expressions we denote with $D_{m,n}$ a field with indices in the Young Tableau representation with two columns, one of length $m$ and one of length $n$.  For instance, this means that the $D_{7,1}$ field has eight indices in total, seven of which are totally antisymmetric and such that antisymmetrising all eight indices one obtains zero.}
\begin{equation}
D_6 \quad  D_{7,1} \quad  D_{8,2} \quad D_{9,3} \quad .
\label{mixedsymmetryNSpotentials}
\end{equation}
More precisely, each potential in \autoref{mixedsymmetryNSpotentials}, after reduction to $D$ dimensions, is dual to each of the fluxes listed in
\autoref{listofNSfluxes}. For instance, the 6-form $D_6$  is dual to the $H$-flux $H_{abc}$ because by reduction one gets a $(D-1)$-form $D_{D-1, a_1... a_{d-3}}$ which is equivalent to $D_{D-1}{}^{abc}$ as a representation of $\text{SL}(d,\mathbb{R})$. The same applies to the other mixed-symmetry fields given in \autoref{mixedsymmetryNSpotentials}. This means that one can split the duality relation \autoref{dualityNSpotentialsDdim} into four different $D$-dimensional relations, one for each flux, as follows:
\begin{eqnarray}
D_6\,:\hskip 1truecm & & \frac{1}{\sqrt{|g|}}\epsilon^{\mu_1 ...\mu_D} \partial_{\mu_1} D_{\mu_2 ..\mu_D}{}^{abc} = {\cal M}^{ad} {\cal M}^{be} {\cal M}^{cf} H_{def} \quad ,\nonumber \\[.1truecm]
D_{7,1}\,:\hskip 1truecm & & \frac{1}{\sqrt{|g|}}\epsilon^{\mu_1 ...\mu_D} \partial_{\mu_1} D_{\mu_2 ..\mu_D}{}^{ab}{}_c = {\cal M}^{ad} {\cal M}^{be} {\cal M}_{cf} f_{de}{}^{f}\quad , \nonumber \\[.1truecm]
D_{8,2}\,:\hskip 1truecm& & \frac{1}{\sqrt{|g|}}\epsilon^{\mu_1 ...\mu_D} \partial_{\mu_1} D_{\mu_2 ..\mu_D}{}^a{}_{bc} = {\cal M}^{ad} {\cal M}_{be} {\cal M}_{cf} Q_{d}{}^{ef}\quad , \nonumber \\[.1truecm]
D_{9,3}\,:\hskip 1truecm & & \frac{1}{\sqrt{|g|}}\epsilon^{\mu_1 ...\mu_D} \partial_{\mu_1} D_{\mu_2 ..\mu_D, abc} = {\cal M}_{ad} {\cal M}_{be} {\cal M}_{cf} R^{def} \quad .\label{NSdualityrelations}
\end{eqnarray}
The last two duality relations  show that the globally non-geometric $Q$ flux is related to a mixed-symmetry tensor with $8$ antisymmetric indices while the locally non-geometric $R$ flux corresponds to a mixed-symmetry tensor with $9$ antisymmetric indices.

%\begin{eqnarray}
%& & \epsilon^{\mu_1 ...\mu_7 abc} \partial_{\mu_1} D_{\mu_2 ..\mu_7} = {\cal M}^{ad} {\cal M}^{be} {\cal M}^{cf} H_{def} \nonumber \\
%& & \epsilon^{\mu_1 ...\mu_8 ab} \partial_{\mu_1} D_{\mu_2 ..\mu_8, c} = {\cal M}^{ad} {\cal M}^{be} {\cal M}_{cf} f_{de}{}^{f} \nonumber \\
%& & \epsilon^{\mu_1 ...\mu_9 a} \partial_{\mu_1} D_{\mu_2 ..\mu_9, bc} = {\cal M}^{ad} {\cal M}_{be} {\cal M}_{cf} Q_{d}{}^{ef} \nonumber \\
%& & \epsilon^{\mu_1 ...\mu_{10}} \partial_{\mu_1} D_{\mu_2 ..\mu_{10}, abc} = {\cal M}_{ad} {\cal M}_{be} {\cal M}_{cf} R^{def} \quad , \label{NSdualityrelations}
%\end{eqnarray}
Although it is not known how to introduce mixed-symmetry potentials at the interacting level into IIA or IIB supergravity, it is nevertheless instructive to think about
how in principle the above  duality relations  could be uplifted to ten dimensions, given that all the fields involved are ten-dimensional fields. For the first relation this is obvious since it does not involve a mixed-symmetry potential. Indeed, because the left-hand side can be uplifted to $ \epsilon^{\mu_1 ...\mu_{10}} \partial_{\mu_1} D_{\mu_2 ..\mu_7}$ and one ends up with the duality relation between $B_2$ and $D_6$ in ten dimensions. The second relation  is only consistent if the lower index $c$ denotes an isometric direction. This means that one gets $\epsilon^{\mu_1 ...\mu_{10}} \partial_{\mu_1} D_{\mu_2 ..\mu_8, c}$ where one of the ten indices $\mu_1 ... \mu_{10}$ are parallel to $c$, but the field does not depend on $x^c$. Similarly, in the other two cases the lower indices $bc$ and $abc$ correspond to isometric directions. In particular, in the last case one ends up with $\epsilon^{\mu_1 ...\mu_{10}} \partial_{\mu_1} D_{\mu_2 ..\mu_{10}, abc}$ which means that also three of the $\mu$ indices of the field must coincide with the isometric indices $abc$.

If one considers the uplift to ten dimensions of the duality relations in the way  discussed above, the $D$ fields always depend on the standard coordinates, with the exception of those corresponding to the isometric directions. The characteristics of the fluxes on the right-hand side of the duality relations is thus mapped to the index properties of the corresponding mixed-symmetry potential. In particular, the global non-geometric nature of the $Q$ flux is translated into the fact that the dual potential is a mixed-symmetry potential with a set of eight antisymmetric indices, while the local non-geometric nature of the $R$ flux  corresponds to the fact that the dual potential is in this case a mixed-symmetry potential with a set of nine antisymmetric indices. We take this as a general rule. The incompatibility of the $R$ flux with the strong constraint in DFT ({\it i.e.}~the fact that it corresponds to a SS reduction in ${\tilde{x}}$) is equivalent to the impossibility of writing a consistent coupling to the mixed-symmetry potential $D_{9,3}$ in ten-dimensional supergravity. The difference with the Romans case is that there one has a 9-form potential $C_9$ that is not a mixed-symmetry field. This implies that $C_9$ is a well defined potential in ten-dimensions, and its field strength is dual to the Romans mass parameter.

The potentials of \autoref{mixedsymmetryNSpotentials} are all contained in the decomposition of the adjoint representation of $\text{E}_{11}$ in mixed-symmetry fields of the ten-dimensional IIA and IIB supergravities \cite{Riccioni:2007au}. This was precisely the result that was used in \cite{Bergshoeff:2011zk} to list these potentials in the context of the classification of the 1/2-BPS solitonic branes in maximal supergravity. These branes have a tension that scales like $g_{S}^{-2}$ in the string frame (where $g_S$ is the string coupling). In \cite{Bergshoeff:2011ee} this was extended to consider the branes with a tension scaling like $g_{S}^{-3}$, and the mixed-symmetry potentials associated to these branes have been classified. Using this as input, we will generalize the the analysis of this section and consider the gaugings that are sourced by domain walls with a tension scaling like $g_{S}^{-3}$. This is the aim of the next section.

\section{\label{newfluxessection}The $P$-fluxes}

In this section we want to extend the analysis performed in \autoref{NSfluxessection} to the so-called $P$-fluxes. As the simplest example of a $P$-flux,
consider a IIB SS reduction to nine dimensions where the field that has a linear dependence on the internal coordinate is the scalar field $\gamma$, which is the S-dual of the axion $C$ and it is defined as the real part of $\tilde{\tau} = - 1/\tau$. In nine dimensions, this flux is a singlet under T-duality, but in lower dimensions it is mapped to other geometric and non-geometric $P$-fluxes.  Another example of a flux belonging to this T-dual family is the S dual of the $Q$ flux discussed in the previous section. A partial classification of these fluxes, with particular attention to those of them that are locally geometric, was obtained in \cite{Aldazabal:2010ef} (see also \cite{Sakatani:2014hba} for an analysis of these fluxes and their relation to branes). A general classification of $P$-fluxes is the subject of this section.

In any dimension, the embedding tensor of the gaugings resulting from the fluxes we are considering in this section belongs to the vector-spinor `gravitino' irreducible representation of $\text{SO}(d,d)$. More precisely, using the conventions for the RR fields such that the RR embedding tensor is $\theta_\alpha$, with $\alpha$ a chiral-spinor index, this embedding tensor is $\theta_{M\alpha}$ for $D$ odd and $\theta_{M \dot{\alpha}}$ for $D$ even.\footnote{This can be explicitly checked by decomposing the representations of the global symmetry groups found in \cite{embeddingtensorreprs} in various dimensions as representations of $\text{SO}(d,d)$.}
The dimension of the representation is $(2d-1) \times 2^{d-1}$. In 9D we have $\text{SO}(1,1)$, and the representation is clearly one-dimensional which implies that this flux is a singlet as just mentioned. In general, to identify all the possible fluxes, one decomposes the gravitino representation in terms of representations
of $\text{SL}(d,\mathbb{R})$, precisely as one does for the NS fluxes that result in the embedding tensor $\theta_{MNP}$ (see the previous section). There are two different decompositions, corresponding to the two possible conventions that one can use for the chiral index $\alpha$ which are given in \autoref{alphaindexIIAconvention} and \autoref{alphaindexIIBconvention}. The IIA case for $d$ odd and the IIB case for $d$ even give
\begin{equation}
( {\bf d} \oplus {\bf \overline{d} }) \otimes \left[  {\bf 1} \oplus {{\bf {d}} \choose { 2}} \oplus {{\bf {d}} \choose {4}} \oplus ... \right] \ominus \left[ {\bf {d}} \oplus  {{\bf {d}} \choose { 3}} \oplus {{\bf {d}} \choose { 5}} \oplus ... \right] \quad ,\label{decompositionofthetaudergldIIADodd}
\end{equation}
while the IIA case for $d$ even and the IIB case for $d$ odd give
\begin{equation}
( {\bf d} \oplus {\bf \overline{d} }) \otimes \left[ {\bf {d}} \oplus  {{\bf {d}} \choose {3}} \oplus {{\bf {d}} \choose { 5}} \oplus ... \right] \ominus \left[  {\bf 1} \oplus {{\bf {d}} \choose { 2}} \oplus {{\bf {d}} \choose { 4}} \oplus ... \right] \quad .\label{decompositionofthetaudergldIIADeven}
\end{equation}
It is understood that in these expressions the representations at the right of the symbol $\ominus$ are subtracted to the ones on its left due to irreducibility.

At first sight, the expressions given above seem to give different representations according to whether $d$ is even or odd for both IIA and IIB, which is inconvenient because we want a unique set of fluxes for each theory. The fact that the set of fluxes indeed is unique  stems from the fact that in $\text{SL}(d,\mathbb{R})$, due to the existence of the $\epsilon$ invariant tensor, the following equivalence between representations holds:
\begin{equation}
{{\bf d} \choose n} \equiv {{\bf \overline{d}} \choose d-n} \quad .
\end{equation}
As a consequence, when one sums over all even-rank or odd-rank antisymmetric representations, for $d$ is even one has the following identities
\begin{eqnarray}
& & {\bf 1} \oplus {{\bf {d}} \choose { 2}} \oplus {{\bf {d}} \choose { 4}} \oplus ... =  {\bf 1} \oplus {{\bf \overline{d}} \choose { 2}} \oplus {{\bf \overline{d}} \choose { 4}} \oplus ...\nonumber \\
& &  {\bf {d}} \oplus  {{\bf {d}} \choose { 3}} \oplus {{\bf {d}} \choose { 5}} \oplus ... =  {\bf \overline{d}} \oplus  {{\bf \overline{d}} \choose { 3}} \oplus {{\bf \overline{d}} \choose { 5}} \oplus ... \quad ,
\end{eqnarray}
while if $d$ is odd one gets the equations
\begin{eqnarray}
& & {\bf 1} \oplus {{\bf {d}} \choose {2}} \oplus {{\bf {d}} \choose {4}} \oplus ... =
{\bf \overline{d}} \oplus  {{\bf \overline{d}} \choose { 3}} \oplus {{\bf \overline{d}} \choose {5}} \oplus ... \nonumber \\
& &  {\bf {d}} \oplus  {{\bf {d}} \choose { 3}} \oplus {{\bf {d}} \choose { 5}} \oplus ... = {\bf 1} \oplus {{\bf \overline{d}} \choose { 2}} \oplus {{\bf \overline{d}} \choose {4}} \oplus ... \quad .
\end{eqnarray}
Substituting these expressions into \autoref{decompositionofthetaudergldIIADodd} and \autoref{decompositionofthetaudergldIIADeven}, one finds a unique expression for each theory. In particular, one gets
\begin{equation}
 ( {\bf d} \oplus {\bf \overline{d} }) \otimes \left[ {\bf \overline{d}} \oplus  {{\bf \overline{d}} \choose { 3}} \oplus {{\bf \overline{d}} \choose { 5}} \oplus ... \right] \ominus \left[  {\bf 1} \oplus {{\bf \overline{d}} \choose { 2}} \oplus {{\bf \overline{d}} \choose { 4}} \oplus ... \right]  \label{embeddingtensorIIAfinaldecomp}
\end{equation}
in the IIA case and
\begin{equation}
 ( {\bf d} \oplus {\bf \overline{d} }) \otimes \left[ {\bf 1} \oplus  {{\bf \overline{d}} \choose { 2}} \oplus {{\bf \overline{d}} \choose { 4}} \oplus ... \right] \ominus \left[  {\bf \overline{d}} \oplus {{\bf \overline{d}} \choose {3}} \oplus {{\bf \overline{d}} \choose {5}} \oplus ... \right] \label{embeddingtensorIIBfinaldecomp}
\end{equation}
in the IIB case.

Writing down the $\text{SL}(d,\mathbb{R})$ representations given in \autoref{embeddingtensorIIAfinaldecomp}  and \autoref{embeddingtensorIIBfinaldecomp} in components, one obtains the fluxes
\begin{eqnarray}
& & P_a^b \quad P_a^{b_1 b_2 b_3} \quad P_a^{b_1 ...b_5} \quad ... \nonumber \\
& & P^{ab} \quad P^{a b_1 b_2 b_3} \quad P^{a b_1 ...b_5} ... \label{thetaincomponentsIIA}
\end{eqnarray}
originating from the IIA theory and
\begin{eqnarray}
& & P_a \quad P_a^{b_1 b_2 } \quad P_a^{b_1 ...b_4} \quad P_a^{b_1 ...b_6}\quad ... \nonumber \\
& & P^{a b_1 b_2} \quad P^{a b_1 b_2 b_3 b_4} \quad P^{a b_1 ...b_6} \quad ... \label{thetaincomponentsIIB}
\end{eqnarray}
originating from the IIB theory. In these expressions, the indices $b_1...b_n$ are completely antisymmetrised, and the representations with all upstairs indices $a b_1 ...b_n$ are irreducible with vanishing completely antisymmetric part. The representations with the $a$ index downstairs and some $b$ indices upstairs are reducible, with the condition that the singlet is always removed. The reader can check that the fluxes given in \autoref{thetaincomponentsIIA} and \autoref{thetaincomponentsIIB} with the aforementioned conditions precisely give the representations listed in \autoref{embeddingtensorIIAfinaldecomp} and \autoref{embeddingtensorIIBfinaldecomp}, respectively. We give in \autoref{RRSdualgaugings} the full list of representations corresponding to the various fluxes as representations of $\text{SL}(d,\mathbb{R})$ in any dimension.

\begin{table}[t!]
\renewcommand{\arraystretch}{1.3}
\par
\begin{center}
\scalebox{.72}{
\begin{tabular}{|c||c||c|c|c|c|c|c||c|c|c|c|c|c|c|}
\hline
$D$ & $\theta_{MA}$ & $P_a^b$  & $P_{a}^{b_1 ...b_3}$ & $P_{a}^{b_1...b_5}$ & $P^{ab}$ & $P^{a b_1 ...b_3}$ & $P^{a b_1 ...b_5}$ & $P_a$ & $P_a^{b_1 b_2}$ & $P_a^{b_1 ...b_4}$  & $P_a^{b_1 ...b_6}$ & $P^{ab_1 b_2}$ & $P^{ab_1 ...b_4}$ & $P^{a b_1 ...b_6}$  \\
\hline
\hline 9 & ${1}$ & & &  &${\bf 1}$ & & & ${\bf 1}$ &&&&&&
 \\
 \hline
8 & ${6}$  & ${\bf 3}$ & &  &${\bf 3}$ & & & ${\bf 2}$ & ${\bf 2}$ & & & ${\bf 2}$ &&\\
\hline
7 & ${20}$  & ${\bf 8}$ & ${\bf 3}$ &  &${\bf \overline{6}}$ & ${\bf \overline{3}}$ & & ${\bf 3}$ & ${\bf 6} \oplus {\bf \overline{3}}$ & & & ${\bf 8}$ &&\\
\hline
6 & ${56}$ & ${\bf 15}$ & ${\bf 10}\oplus {\bf {6}}$ &  &${\bf \overline{10}}$ & ${\bf 15}$ & & ${\bf 4}$ & ${\bf 20} \oplus {\bf \overline{4}}$ & ${\bf 4}$& & ${\bf \overline{20}}$ &${\bf \overline{4}}$ &    \\
\hline
5 & ${144}$ & ${\bf 24}$ & ${\bf 40}\oplus {\bf \overline{10}}$ & ${\bf 5}$ &${\bf \overline{15}}$ & ${\bf \overline{45}}$ & ${\bf \overline{5}}$ & ${\bf 5}$ & ${\bf 45} \oplus {\bf \overline{5}}$ & ${\bf 15}\oplus {\bf 10}$& & ${\bf \overline{40}}$ &${\bf 24}$ &  \\ \hline
4 & ${352}$ & ${\bf 35}$ & ${\bf 105}\oplus {\bf \overline{15}}$ & ${\bf 21}\oplus {\bf 15}$ &${\bf \overline{21}}$ & ${\bf \overline{105}}$ & ${\bf 35}$ & ${\bf 6}$ & ${\bf 84} \oplus {\bf \overline{6}}$ & ${\bf 70}\oplus {\bf 20}$& ${\bf 6}$ & ${\bf \overline{70}}$ &${\bf \overline{84}}$ & ${\bf \overline{6}}$   \\
\hline
\end{tabular}
}
\end{center}
\caption{{\footnotesize In this Table we give the decomposition in terms of $\text{SL}(d,\mathbb{R})$ representations of the $\theta_{MA}$ embedding tensor for both the IIA case (left of double vertical line in the middle) and the IIB case (right of double vertical line in the middle). The embedding tensor $\theta_{MA}$ is given by $\theta_{M\alpha}$ for $D$ odd and by $\theta_{M\dot\alpha}$ for $D$ even.}}
\label{RRSdualgaugings}
\end{table}

The general expressions given in \autoref{embeddingtensorIIAfinaldecomp} and \autoref{embeddingtensorIIBfinaldecomp} show that all these gaugings can be obtained as SS reductions in double space for the fields that are the T-duals of the field $\gamma$ mentioned at the beginning of this section. These fields are
\begin{equation}
\gamma^a \qquad \gamma^{a_1 ...a_3} \qquad \gamma^{a_1 ...a_5} \ ...
\end{equation}
in the case of IIA supergravity and
\begin{equation}
\gamma \qquad \gamma^{a_1 a_2} \qquad \gamma^{a_1 ...a_4} \ ...
\end{equation}
in the caee of IIB supergravity \cite{Aldazabal:2010ef}.
In particular, the geometric gaugings are
  \begin{equation}
P_a^{b_1 ... b_n} = \partial_a \gamma^{b_1 ... b_m} \quad ,
  \end{equation}
while the non-geometric ones are
\begin{equation}
 P^{a b_1 ...b_n} = \tilde{\partial}^a \gamma^{b_1 ... b_n} - \tilde{\partial}^{[a} \gamma^{b_1 ... b_n ]} \quad .
\end{equation}
% In $\text{SO}(d,d)$-covariant notation, the RR scalars are $C_{\dot{\alpha}}$, and their S-duals are $\gamma_{\dot{\alpha}}$ for $D$ even and
% $\gamma_{{\alpha}}$ for $D$ odd. In general we write the SS ansatz as
%   \begin{equation}
%   \theta_{M A} = \partial_M \gamma_A  \quad ,
%   \end{equation}
% where $A$ is either $\alpha$ ($D$ odd) or $\dot{\alpha}$ ($D$ even). The irreducibility condition on the embedding tensor, $(\Gamma^M )_A{}^B \theta_{MB} % =0$, results in the condition $\slashed{\partial} \gamma =0$.

The analysis that we have just performed shows that all the fluxes we are considering admit a realisation as generalised SS fluxes in double field theory. We have made a distinction  between the fluxes with a lower index, that one  expects to be locally geometric \cite{Aldazabal:2010ef} and for which the strong constraint should not be violated, and the fluxes with all upstairs indices, that are non-geometric and do not satisfy the strong constraint. In the previous section we have shown that the NS fluxes are dual to $D-1$-forms coming from mixed-symmetry potentials in ten dimensions, and we have shown that the non-geometric $R$ flux is dual to the potential $D_{9,3}$ with a set of nine antisymmetric indices. We will now determine all the potentials that are dual to the fluxes listed in \autoref{thetaincomponentsIIA} and \autoref{thetaincomponentsIIB}. We will show that also in this case the non-geometric fluxes are dual to potentials with a set of nine antisymmetric indices.

The $D-1$-form field that is dual to the embedding tensor we are discussing in this section is the $D-1$-form $E_{D-1, M \dot{\alpha}}$. Neglecting the contribution from the other fields the duality relation, which is the analogue of the NS duality relation \eqref{dualityNSpotentialsDdim} in  the previous section,  reads
\begin{equation}
\frac{1}{\sqrt{|g|}}\epsilon^{\mu_1 ...\mu_D} \partial_{\mu_1} E_{\mu_2 ..\mu_D, M \dot{\alpha}} = {\cal M}_M{}^N  C_{\dot{\alpha} \dot{\beta}} \mathbb{S}^{\dot{\beta} \dot{\gamma}} \theta_{N \dot{\gamma}}
\end{equation}
for $d$ even and
\begin{equation}
\frac{1}{\sqrt{|g|}}\epsilon^{\mu_1 ...\mu_D} \partial_{\mu_1} E_{\mu_2 ..\mu_D, M \dot{\alpha}} = {\cal M}_M{}^N  C_{\dot{\alpha} {\beta}} \mathbb{S}^{{\beta} {\gamma}} \theta_{N {\gamma}} \label{dualityrelationED-1theta}
\end{equation}
for $d$ odd. Here $C_{\dot{\alpha} \dot{\beta}}$ and $C_{\dot{\alpha} {\beta}}$ are the $\text{O}(d,d)$ charge conjugation matrices for $d$ even and $d$ odd, respectively.
  The matrices $\mathbb{S}^{\dot{\alpha} \dot{\beta}}$ and $\mathbb{S}^{{\alpha} {\beta}}$ in the above duality relations are the equivalent of ${\cal M}$ in the spinor representation. Note that the chirality properties of the $\text{O}(d,d)$ charge conjugation matrices are precisely the ones that are needed to write down  duality relations that are consistent with the chirality properties of the potentials and fluxes.

The field $E_{D-1, M \dot{\alpha}}$ arises  in any dimension from the reduction of the ten-dimensional mixed-symmetry fields
\begin{equation}
E_{8,1} \quad E_{8,3} \quad E_{8,5} \quad E_{9,1,1} \quad E_{9,3,1} \quad E_{9,5,1} \label{mixedsymmetryEpotentialsIIA}
\end{equation}
in the IIA case and
\begin{equation}
E_8 \quad E_{8,2} \quad E_{8,4} \quad E_{8,6} \quad E_{9,2,1} \quad E_{9,4,1} \quad E_{9,6,1} \label{mixedsymmetryEpotentialsIIB}
\end{equation}
in the IIB case. These potentials have been already listed in \cite{Bergshoeff:2011ee} in the context of the classification of the 1/2-BPS branes whose tension scales like $g_S^{-3}$, and here we are using exactly the notation of that paper. By explicitly performing the dimensional reduction, one can show how these fields precisely build the gravitino representation for the $D-1$ form in $D$ dimensions. The result is summarised in \autoref{mixedsymmetrypotentialsdecomp}.

\begin{table}[t!]
\renewcommand{\arraystretch}{1.3}
\par
\begin{center}
\scalebox{.75}{
\begin{tabular}{|c||c||c|c|c|c|c|c||c|c|c|c|c|c|c|}
\hline
$D$ & $E_{M\dot{\alpha}}$ & $E_{8,1}$  & $E_{8,3}$ & $E_{8,5}$ & $E_{9,1,1}$ & $E_{9,3,1}$ & $E_{9,5,1}$ & $E_8$ & $E_{8,2}$ & $E_{8,4}$  & $E_{8,6}$ & $E_{9,2,1}$ & $E_{9,4,1}$ & $E_{9,6,1}$  \\
\hline
\hline 9 & ${1}$ & & &  &${\bf 1}$ & & & ${\bf 1}$ &&&&&&
 \\
 \hline
8 & ${6}$  & ${\bf 3}$ & &  &${\bf 3}$ & & & ${\bf 2}$ & ${\bf 2}$ & & & ${\bf 2}$ &&\\
\hline
7 & ${20}$  & ${\bf 8}$ & ${\bf \overline{3}}$ &  &${\bf {6}}$ & ${\bf {3}}$ & & ${\bf \overline{3}}$ & ${\bf \overline{6}} \oplus {\bf {3}}$ & & & ${\bf 8}$ &&\\
\hline
6 & ${56}$ & ${\bf 15}$ & ${\bf \overline{10}}\oplus {\bf \overline{6}}$ &  &${\bf {10}}$ & ${\bf 15}$ & & ${\bf \overline{4}}$ & ${\bf \overline{20}} \oplus {\bf {4}}$ & ${\bf \overline{4}}$& & ${\bf {20}}$ &${\bf {4}}$ &    \\
\hline
5 & ${144}$ & ${\bf 24}$ & ${\bf \overline{40}}\oplus {\bf {10}}$ & ${\bf \overline{5}}$ &${\bf {15}}$ & ${\bf {45}}$ & ${\bf {5}}$ & ${\bf \overline{5}}$ & ${\bf \overline{45}} \oplus {\bf {5}}$ & ${\bf \overline{15}}\oplus {\bf \overline{10}}$& & ${\bf {40}}$ &${\bf 24}$ &  \\ \hline
4 & ${352}$ & ${\bf 35}$ & ${\bf \overline{105}}\oplus {\bf {15}}$ & ${\bf \overline{21}}\oplus {\bf \overline{15}}$ &${\bf {21}}$ & ${\bf {105}}$ & ${\bf 35}$ & ${\bf \overline{6}}$ & ${\bf \overline{84}} \oplus {\bf {6}}$ & ${\bf \overline{70}}\oplus {\bf 20}$& ${\bf \overline{6}}$ & ${\bf {70}}$ &${\bf {84}}$ & ${\bf {6}}$   \\
\hline
\end{tabular}
}
\end{center}
\caption{{\footnotesize In this Table we give the $\text{SL}(d,\mathbb{R})$ representations that build up the vector-spinor representation $E_{M\dot\alpha}$ for both the IIA case (left of double vertical line in the middle) and the IIB case (right of double vertical line in the middle).}}
\label{mixedsymmetrypotentialsdecomp}
\end{table}

By comparing \autoref{RRSdualgaugings} and \autoref{mixedsymmetrypotentialsdecomp} it is clear that the geometric fluxes, that is the ones of the form $P_a^{b_1 ...b_n}$, are dual to the mixed-symmetry potentials with a set of 8 antisymmetric indices, while the non-geometric fluxes, of the form $P^{a b_1 ...b_n}$, are dual to the mixed-symmetry potentials with 9 antisymmetric indices. This can be seen explicitly by decomposing the duality relation in \autoref{dualityrelationED-1theta} in terms of the fluxes, which leads to
\begin{eqnarray}
& & \frac{1}{\sqrt{|g|}}\epsilon^{\mu_1 ...\mu_D} \partial_{\mu_1} E_{\mu_2 ..\mu_D}{}^{a}{}_{b_1 ...b_n} = {\cal M}^{ac} {\cal M}_{b_1 d_1} ... {\cal M}_{b_n d_n} P_c^{d_1 ...d_n} \nonumber \\[.1truecm]
& & \frac{1}{\sqrt{|g|}}\epsilon^{\mu_1 ...\mu_D} \partial_{\mu_1} E_{\mu_2 ..\mu_D, a, b_1 ...b_n}  = {\cal M}_{ac} {\cal M}_{b_1 d_1} ...{\cal M}_{b_n d_n} P^{c, d_1 ...d_n }\quad ,\label{Pfluxdualityrelations}
\end{eqnarray}
where $n$ is odd for IIA and even for IIB. The first duality relation involves the locally-geometric fluxes, while the second the non-geometric ones. As in the previous section, these relations can formally be uplifted to ten dimensions, keeping in mind that the lower $b$ and $a$ indices have to be treated as isometric, {\it i.e.}~the field does not depend on the corresponding coordinates. With this restriction taken into account, the potentials on the left-hand side only depend on the standard coordinates, while the $\gamma$ fields generating the fluxes on the right-hand side depend on $x$ or $\tilde{x}$ according to whether the flux is locally geometric or non-geometric.

To summarise, we have shown that all the $P$ fluxes can be realised as generalised SS reductions in double space, implying that all these fluxes must admit a description in DFT.
Exactly as for the NS fluxes, we have also shown that  the geometric $P$ fluxes are dual to mixed-symmetry potentials with 8 antisymmetric indices, while the non-geometric ones are dual to mixed-symmetry potentials with 9 antisymmetric indices.

\section{\label{conclusionssection}Conclusions}

In this paper we have considered how the duality relations between $D-1$-forms and the embedding tensor in any dimension can be rewritten in terms of mixed-symmetry potentials on one side and generalised fluxes on the other. In particular, we have considered the NS fluxes and the $P$ fluxes, and we have shown that the locally geometric fluxes are dual to mixed-symmetry potentials with 8 antisymmetric indices, while the non-geometric fluxes are dual to mixed-symmetry potentials with 9 antisymmetric indices. In these relations, the mixed-symmetry potentials depend on the normal coordinates, and the non-geometric nature of the flux translates to the impossibility of coupling consistently the potential in supergravity.

The $P$ fluxes have a natural characterisation as SS reductions in double space. It would   be interesting to extend DFT in order to include these fluxes. In particular, the $\gamma$ potentials group together to form a spinor representation of $\text{SO}(10,10)$, but the fact that these fields are related to the other fields of the theory might be difficult to implement in DFT. In particular, the S-duality of IIB supergravity, which allows to define the nine-dimensional $P$ flux that was the starting point of our analysis in \autoref{newfluxessection}, also exchanges $B_2$  and $C_2$, and it is therefore not manifest in DFT.

It would be worth studying whether the duality relations between mixed-symmetry potentials and fluxes admit a formulation in DFT. The potentials listed in \autoref{mixedsymmetryNSpotentials} originate from a DFT field $D_{MNPQ}$ with four antisymmetric indices of $\text{SO}(10,10)$, as results from decomposing the $\text{E}_{11}$ Kac-Moody algebra in representations of  $\text{SO}(10,10)$. There should be a way to define such a  field to be the dual to the generalised metric ${\cal H}_{MN}$ of DFT.\,\footnote{One can write down a first-order formulation of the DFT action at the linearized level and in this way derive  a dual formulation
\cite{inprogress}. It is, however, not clear that this leads to the  duality relation discussed in the text.}
Similarly, the potentials listed in \autoref{mixedsymmetryEpotentialsIIA} and \autoref{mixedsymmetryEpotentialsIIB} arise from the $\text{SO}(10,10)$ field $E_{MN, \dot{\alpha}}$, in the tensor-spinor irreducible representation, and one expects this field to be dual to the RR spinor of DFT.

These duality relations, if they  can be formulated in some way, can also be seen as the origin of the existence of the so-called `wrapping rules' \cite{Bergshoeff:2011mh,Bergshoeff:2011ee} satisfied by the branes that are electrically charged with respect to the various potentials of the supergravity theories. For the fundamental branes, the wrapping rule  states that the fundamental string always sees a doubled cycle when it wraps. The branes electrically charged under the $D$ potentials, that we call the $\alpha=-2$ branes (where $T\sim g_S^\alpha$ is the tension of the brane in the string frame) contain the duals of the fundamental branes, and for these the wrapping rule is  the dual, {\it i.e.}~they double when they do not wrap a cycle. The fact that this wrapping rule extends to the $\alpha=-2$ branes that are not dual to fundamental branes would be a natural consequence of a DFT duality relation, that reduces to the standard duality relations only if projected in normal space. Similarly,  the $\alpha=-3$ branes, that are electrically charged under the $E$ fields, always double upon dimensional reduction, regardless of whether they wrap or do not wrap a cycle. These branes are dual to the D-branes, that always see a standard geometry, {\it i.e.}~they never double, and therefore the wrapping rules of the $\alpha=-3$ branes would be a straightforward consequence of the DFT duality relation, if it existed.

As a natural extension, one can consider how the analysis performed in this work can be extended to fluxes that are sourced by domain walls with even more negative values of $\alpha$, that are more and more non-perturbative in string theory. All such domain walls,  and their corresponding mixed-symmetry potentials,   have been classified  \cite{Bergshoeff:2012pm}. It would also be interesting to study how this analysis is generalised to theories with less supersymmetry and reductions on non-flat manifolds. This is crucial if one wants to understand what the presence of these fluxes can teach us.  In particular, the $P$ flux that is the S-dual of the $Q$ flux has been considered recently \cite{Danielsson:2015rca} in a more phenomenological context.

The take-home message of this paper is that we have established a relation between two rather different research activities: non-geometric fluxes and DFT on the one side and mixed-symmetry potentials and supergravity on the other side. Both have their own issues. Non-geometric fluxes can be understood to result from a generalized SS reduction in DFT but the extra dependence on the winding coordinates, which is necessary for the SS reduction, violates the strong constraint. For an attempt to give a geometrical description of such non-geometric fluxes, see \cite{GaugedDFT, Hull:2009sg}. On the other hand, the issue with mixed-symmetry potentials is that we only know how to describe then at the linearized level. Nevertheless, their existence is predicted by $\text{E}_{11}$ \cite{West:2001as}  and mixed-symmetry potentials are expected to play a role in constructing stringy extensions of supergravity. In short, our work suggests that a mild violation of the strong constraint, needed for a SS reduction, is
equivalent to an extension of supergravity involving mixed-symmetry potentials. We hope that this interrelationship might stimulate new developments in both the DFT and  the supergravity approach.

\vskip 1cm

\section*{Acknowledgements}
We are grateful to Tomas Ort\'{\i}n, who was originally involved in the collaboration, for several stimulating discussions and insights.
E.B. and F.R. ~wish to thank the KITP Institute in Beijing, where this work was started,
for  its hospitality and its generous financial support, and the organizers of the
`Quantum Gravity, Black Holes and Strings' program for creating a stimulating atmosphere.
F.R. would like to thank the University of Groningen and E.B. and F.R. would like to thank the IFT in Madrid for the kind hospitality at the final stages of this work. V.P. would like to thank Olaf Hohm, Diego Marques and Giuseppe Dibitetto for useful comments.

\end{document}